\documentclass[aps,pra,twocolumn,showpacs]{revtex4}
\usepackage{graphicx}
\usepackage{color} 
\usepackage{ulem} 
\usepackage{amsmath,amssymb}

\begin{document}

\title{Quantification of Nonclassicality}

\author{C. Gehrke}\affiliation{Arbeitsgruppe Theoretische Quantenoptik, Institut f\"ur Physik, Universit\"at Rostock, D-18051 Rostock, Germany}
\author{J. Sperling}\affiliation{Arbeitsgruppe Theoretische Quantenoptik, Institut f\"ur Physik, Universit\"at Rostock, D-18051 Rostock, Germany}
\author{W. Vogel}\affiliation{Arbeitsgruppe Theoretische Quantenoptik, Institut f\"ur Physik, Universit\"at Rostock, D-18051 Rostock, Germany}

\begin{abstract}
To quantify single mode nonclassicality, we start from an operational approach. A positive semi-definite observable is introduced to describe a measurement setup. The quantification is based on the negativity of the normally ordered version
of this observable. Perfect operational quantumness corresponds to the quantum-noise-free measurement of the chosen observable. Surprisingly, even moderately squeezed states may exhibit perfect quantumness for a properly designed measurement.
The quantification is also considered from an axiomatic viewpoint, based on the algebraic structure of the quantum states and the quantum superposition principle. Basic conclusions from both approaches are consistent with this fundamental principle of the quantum world.
\end{abstract}


\maketitle

%============================================================================================================================================================================
\section{Introduction}

Experimental realizations of nonclassical effects of light opened 
interesting perspectives for practical applications of nonclassical quantum states. Consequently, nonclassical states of light and matter have attracted substantial interest during the last decades. In this context, the quantitative characterization of nonclassical effects is an important issue. From an operational point of view, it is of some interest to connect a quantitative characterization of quantum effects with its potential applications. This includes the suppression of quantum noise in different types of measurements. From the fundamental point of view, the quantification should be related to fundamental principles of quantum physics.

There exists a number of attempts to quantify the nonclassicality of a harmonic oscillator quantum system. Among them, Hillery introduced the concept of the distance between two quantum states~\cite{pra-35-725}. He defined the distance from the classical states as a quantitative measure of nonclassicality. Although this is an intuitive approach, in many cases the nonclassical distance is hard to calculate. 
Another measure of nonclassicality was introduced by Lee \cite{pra-44-R2775}, the nonclassical depth of a quantum state. It is defined by the minimum number of thermal photons admixed to a quantum state, which is needed to destroy its nonclassical effects. This quantity, however, 
is essentially a measure of the fragility of quantum effects under certain thermal disturbances.
Asb\'{o}th et~al. consider the amount of entanglement, which can be potentially generated by splitting a nonclassical state by a beam splitter, as a measure of nonclassicality~\cite{prl-94-173602}. 
Despite interesting relations between nonclassicality and entanglement~\cite{pra-65-032323,pra-66-024303}, this 
measure is based on a special class of all quantum effects.
Moreover, an entanglement potential suffers from the difficulty to define a general entanglement measure, cf. e.g.~\cite{horodecki,guehne}.

In Quantum Optics, nonclassicality of a quantum state of the harmonic oscillator is characterized by negativities of the Glauber-Sudarshan $P$~function~\cite{Sudarshan,Glauber,pr-140-B676}, which belongs to the class of quasiprobabilities. Our further studies are based on this definition of nonclassicality. For many quantum states the $P$~function has not only negativities, it can also be strongly singular. Recently the concept of nonclassicality quasiprobabilities has been developed. The latter are regularized versions of the $P$~function, which are accessible in experiments and completely identify all negativities of the $P$~function~\cite{pra-82-032107,pra-83-032116,prl-107-113604}.

Extending the $P$~function to the multi-mode case, its negativities also include entanglement as a special nonclassical effect. Note that negativities of the $P$~function are necessary for entanglement but not sufficient. For an unambiguous identification of entanglement, entanglement quasiprobabilities have been introduced~\cite{pra-82-042337,njp-14-055026}. They are also regular functions and their negativities are necessary and sufficient for the existence of any kind of entanglement.
These two concepts, nonclassicality and entanglement quasiprobabilities, allow one to describe both phenomena on a unified footing.

In the present paper we will introduce an operational quantification of nonclassicality in terms of 
experimentally accessible quantities. This will lead to conditions which are directly related to the amount of quantum noise in specific measurement scenarios. The other way around, based on this definition of the amount of nonclassicality, one may calculate the structure of the quantum states which are best suited for the suppression of quantum noise in a certain experimental setup. We also consider the quantification of nonclassicality from a more general, axiomatic point of view and compare it with the operational approach.

The paper is organized as follows. In Section~\ref{sec:quantif} we introduce the  operational quantification of nonclassicality. Established examples, such as quadrature squeezing and sub-Poisson number statistics, are considered in~Section~\ref{sec:establ-ex} from the perspective of our approach. 
The practical relevance of our operational quantification is studied in Section~\ref{sec:sq-nfm} in the framework of measurements free of quantum noise. 
In Section~\ref{sec:gen-quant} the nonclassicality quantification is studied from a general axiomatic point of view and the conclusions drawn from this perspective are compared with the operational approach.
A summary and some conclusions are given in Section~\ref{sec:sum}.

%============================================================================================================================================================================
\section{Operational quantification}
\label{sec:quantif}

In this section we will introduce operational measures for nonclassicality or quantumness,
by starting from the established notion of nonclassicality for quantum states of harmonic oscillators. These measures will be directly based on observable mean values as they are obtained by a chosen experimental setup. This is just what an experimenter needs for a certain application. Either some source
of nonclassical states is given and an optimal measurement technique is sought, or, the other way around,
for a given detection device on may seek for the quantum state optimizing the system operation. 
Our approach yields a systematic method for implementing quantum-noise-free (QNF) measurement techniques.

Let us start with a reformulated definition of nonclassicality, which is equivalent to the failure of the $P$~function to be a probability distribution. It is based on the negativity 
of expectation values whose classical counterparts are positive semi-definite. A quantum state is nonclassical, if there exists an observable $\hat{f}^\dagger \hat{f}$, with  $\hat{f}\equiv
\hat{f}(\hat{a},\hat{a}^\dagger )$ being an operator function of the annihilation (creation) operator $\hat{a}$ ($\hat{a}^\dagger$), so that \cite{pra-71-011802,prl-94-153601,pra-82-013842}
\begin{equation}\label{eq:noncl-cond}
	\langle : \hat{f}^\dagger \hat{f} : \rangle < 0\,,
\end{equation}
where the '':\,\dots:'' symbol denotes normal ordering. 

To introduce an operational measure of nonclassicality, let us consider an
experimental setup which is characterized by an arbitrary but fixed operator $\hat{f}$. The resulting quantities $\hat{f}^\dagger \hat{f}$ and
$:\hat{f}^\dagger \hat{f}:$ are Hermitian operators and hence observables of the chosen setup. Whereas the first observable is positive semi-definite, the second one may have negativities in its spectrum. For the following we assume 
that the nonclassicality Condition~(\ref{eq:noncl-cond}) is fulfilled
for the chosen operator $\hat{f}$ and the quantum state under study. Only in this case the state is identified as being nonclassical and the quantification is meaningful. 

To quantify the nonclassicality of a given state in a certain experimental context, we attempt to properly quantify the negativity that can
be attained by the left-hand side (lhs) of the Condition~(\ref{eq:noncl-cond}). 
Let us consider the difference $\Delta$ between the normally ordered and the ordinary expectation values of the chosen observable 
$\hat{f}^\dagger \hat{f}$,
\begin{equation}\label{eq:Delta}
\Delta =\langle : \hat{f}^\dagger \hat{f} : \rangle - \langle  \hat{f}^\dagger \hat{f}  \rangle\,.
\end{equation}
For a given operator $\hat{f}$ it is straightforward to derive an explicit expression of the quantity $\Delta$ by methods of operator ordering, but this is not needed for the following considerations.
Since $\langle \hat{f}^\dagger \hat{f} \rangle \ge 0 $, the relation 
\begin{equation}\label{eq:min}
\Delta \le \langle : \hat{f}^\dagger \hat{f} : \rangle <0 
\end{equation}
holds true.
 
Now we may define the operational relative nonclassicality $R$ of a given
quantum state for a chosen measurement scheme as
\begin{equation}\label{eq:ratio}
	R=\left\lbrace\begin{array}{cl}
		\frac{\langle : \hat{f}^\dagger \hat{f} : \rangle}{\Delta} & \text{for}\quad \langle : \hat{f}^\dagger \hat{f} : \rangle<0,
		\\ 0 & \text{else}.
	\end{array}\right.
\end{equation}
This ratio quantifies the negativity of the lhs of the Condition~\eqref{eq:noncl-cond} relative to the lower bound set by the Condition~(\ref{eq:min}). 
Based on this definition, a quantum state exhibits perfect nonclassicality, that is $R=1$, if the (negative) value of $\langle : \hat{f}^\dagger \hat{f} : \rangle$ approaches the corresponding lower bound. Hence, perfect nonclassicality as defined on this basis is attained for 
\begin{equation}\label{eq:Delta-nf}
\Delta =   \langle : \hat{f}^\dagger \hat{f} : \rangle \quad
\Longleftrightarrow  \quad \langle \hat{f}^\dagger \hat{f} \rangle =0\,.
\end{equation}
In addition, we have defined $R\equiv0$ for $\langle : \hat{f}^\dagger \hat{f} : \rangle \ge 0$.

Due to the equivalence in Eq.~\eqref{eq:Delta-nf}, perfect quantumness may also be defined by the condition $\langle \hat{f}^\dagger \hat{f} \rangle =0$.
For a general mixed quantum state, described by the density operator $\hat{\varrho} = \sum_\psi
p_\psi | \psi \rangle \langle \psi|$, with $p_\psi > 0$ and $\sum_\psi p_\psi=1$, perfect quantumness requires that 
\begin{equation}\label{eq:perf-qu-mix}
\langle \hat{f}^\dagger \hat{f} \rangle = \sum_\psi p_\psi \|\hat{f}
|\psi\rangle \|^2  =0\,.
\end{equation} 
This condition is fulfilled if and only if 
\begin{equation}\label{eq:perf-qu-state}
\hat{f} | \psi \rangle = 0,
\end{equation} 
for all states $|\psi\rangle$
contained in $\hat{\varrho}$. Thus perfect quantumness is attained for any quantum state composed of eigenstates of the operator $\hat{f}$ whose eigenvalues are zero.
In such cases the observable $\hat{f}^\dagger \hat{f}$ is totally free of quantum noise. 

So far a quantum-noise-free observable $\hat{f}^\dagger \hat{f}$ can only attain the minimal expectation value of zero. This restriction can be easily relaxed. 
In general one can 
substitute $\hat{f} \mapsto \Delta \hat{f} = \hat{f}-\langle \hat{f} \rangle$. Now the condition   
\begin{equation}\label{eq:perf-qu-state-delta}
\Delta \hat{f} |\tilde \psi\rangle =0
\end{equation}
replaces Eq.~\eqref{eq:perf-qu-state}. 
Based on this modified condition we arrive at
\begin{equation}\label{eq:modif}
\langle\tilde\psi | \hat{f}^\dagger \hat{f} | \tilde\psi \rangle = |\langle \tilde\psi | \hat{f} |\tilde\psi\rangle|^2,
\end{equation}
so that the positive semi-definite operator $\hat{f}^\dagger \hat{f}$ can attain nonzero expectation values. Its variance is readily calculated to be
\begin{equation}\label{eq:modif-var}
\langle \tilde\psi | [\Delta(\hat{f}^\dagger \hat{f})]^2  |\tilde\psi\rangle = \langle \tilde\psi |[\hat{f}, \hat{f}^\dagger] |\tilde\psi\rangle  |\langle \tilde\psi | \hat{f} |\tilde\psi\rangle|^2
\end{equation}
Hence, for the choice of a Hermitian operator, $\hat{f} = \hat{f}^\dagger$, the operator $\hat{f}^\dagger
\hat{f}$ has a nonzero mean value and is free of quantum noise, $\langle \tilde\psi | [\Delta(\hat{f}^\dagger \hat{f})]^2  |\tilde\psi\rangle = 0$.

%============================================================================================================================================================================
\section{Elementary quantum effects}
\label{sec:establ-ex}

\subsection{Sub-Poisson number statistics}

Let us start with a well-known nonclassical effect, the sub-Poisson statistics of photons. 
When a radiation field with sub-Poisson photon statistics is recorded by a photodetector, 
the statistics of the photoelectrons also becomes a sub-Poisson one. 
The noise level of the recorded signal can go below the classical shot-noise limit, which is the lowest noise level for the detection with classical light, see e.g.~\cite{book,Ma-Wo-book}. For the first experimental demonstrations of sub-Poisson light we refer to~\cite{prl51-384,josab2-275}.

A qualitative characterization of nonclassicality of sub-Poisson radiation by the Condition~(\ref{eq:noncl-cond}) is rather simple. Let us choose 
\begin{equation}
\label{eq:f-JC}
 \hat{f} \equiv \Delta \hat{n} = \hat{n} - \langle \hat{n} \rangle, 
\end{equation}
with $\hat{n}= \hat{a}^\dagger \hat{a}$ being the photon number operator.
Thus the nonclassicality condition reads as
\begin{equation}\label{eq:noncl-sub-no}
	\langle{:}(\Delta \hat{n})^2{:}\rangle < 0.
\end{equation} 
The classical counterpart of this condition yields $\langle  (\Delta {n})^2  \rangle_{\rm cl}  \ge 0$. 
Applying the fundamental commutation relation, $[\hat{a},\hat{a}^\dagger]=1$, the condition reads as
\begin{equation}\label{eq:noncl-sub}
\langle  (\Delta \hat{n})^2  \rangle <  \langle  \hat{n}  \rangle \,.
\end{equation} 
This shows more clearly that the variance of the photon number is below its value for Poisson statistics. The latter is often referred to as the shot-noise limit.

Let us now consider the operational quantification of the sub-Poisson photon statistics.
The relative nonclassicality defined by Eq.~\eqref{eq:ratio} is of the form
\begin{equation}\label{eq:ratio-sub}
R= -\frac{\langle : (\Delta \hat{n})^2 : \rangle}{\langle \hat{n} \rangle } = 1-\frac{\langle (\Delta \hat{n})^2 \rangle}{\langle \hat{n} \rangle }.
\end{equation}
This result is again intuitively clear. The nonclassical effect attains its maximum value of $R=1$ if the 
variance of the photon number becomes zero, so that the photon number is precisely defined. This is the case when the quantum system is prepared in an eigenstate of the photon number operator.
Note that our general approach of operational nonclassicality quantification   leads for this example  to $R=-Q$, which is the negative value of the Mandel $Q$~parameter~\cite{ol4-205}. The latter has been frequently used for the quantification of the nonclassicality of sub-Poisson light. 

\subsection{Quadrature squeezing}

Now we consider a homodyne detection device for the case of perfect detectors. 
The setup measures the probability distribution $p(x,\varphi)$, of the phase-sensitive quadrature operator,
\begin{equation}\label{eq:modif-qu}
\hat{x}_\varphi=\hat{a} e^{i\varphi} + \hat{a}^\dagger e^{-i\varphi}, 
\end{equation}
of a given radiation mode, cf. e.g.~\cite{book,Ma-Wo-book}.
By choosing  
\begin{equation}\label{eq:noncl-quadr}
\hat{f} \equiv \Delta \hat{x}_\varphi=\hat{x}_\varphi - \langle
\hat{x}_\varphi \rangle\,,
\end{equation}
the Condition~\eqref{eq:noncl-cond} characterizes the effect of quadrature squeezing, 
\begin{equation}\label{eq:noncl-squeeze}
\langle : (\Delta \hat{x}_\varphi)^2 : \rangle < 0\,.
\end{equation}
The classical counterpart of the lhs is $\langle (\Delta {x}_\varphi)^2 \rangle_{\rm cl}$. It represents the classical variance of the stochastic variable ${x}_\varphi$, which is non-negative in general. Quadrature squeezing indicates a reduction of the noise below the vacuum level. For the first experimental demonstrations of this effect we refer to~\cite{prl55-2409,prl57-2520}.   

Let us consider the quantification of the squeezing effect based on the observable $\hat{f}^\dagger \hat{f}$ together with the choice of $\hat{f}$ given by Eq.~\eqref{eq:noncl-quadr}. For this purpose we consider the denominator in Eq.~\eqref{eq:ratio}, which now reads as
\begin{equation}\label{eq:den-squeeze}
\langle : (\Delta \hat{x}_\varphi)^2 : \rangle - \langle  (\Delta \hat{x}_\varphi)^2 \rangle = - \langle  (\Delta \hat{x}_\varphi)^2\rangle_{\rm vac}.
\end{equation}
Here we have used the fact that the quadrature variance differs form its normal-ordered value by the quadrature variance in the vacuum state, $\langle  (\Delta \hat{x}_\varphi)^2\rangle_{\rm vac}$.
Perfect operational nonclassicality for a quadrature measurement of squeezing, i.e. $R=1$ according to Eq.~\eqref{eq:ratio}, requires that
\begin{equation}\label{eq:max-squeeze}
\langle : (\Delta \hat{x}_\varphi)^2 : \rangle  = - \langle  (\Delta \hat{x}_\varphi)^2\rangle_{\rm vac}
\,\, \Longleftrightarrow \,\, \langle  (\Delta \hat{x}_\varphi)^2 \rangle =0,
\end{equation}
see also Eq.~\eqref{eq:Delta-nf}. This is the well-known situation of perfect squeezing, representing the perfect suppression of the vacuum noise for a chosen value of the phase $\varphi$.

The quantum state realizing the so-defined perfect quantumness is a quadrature eigenstate. Such states, however, are unphysical ones since they contain an infinite amount of energy. Squeezed states have some minimal, non-zero value of the quadrature variance, $\langle  (\Delta \hat{x}_\varphi)^2 \rangle= \langle  (\Delta \hat{x}_\varphi)^2 \rangle_{\rm min}$. Their maximal relative nonclassicality, $R_{\rm max}$, is
\begin{equation}\label{eq:ratio-squeeze}
R_{\rm max}= 1 -  \frac{\langle  (\Delta \hat{x}_\varphi)^2 \rangle_{\rm min}}{\langle  (\Delta \hat{x}_\varphi)^2\rangle_{\rm vac}},
\end{equation}
in agreement with an intuitive quantification of quadrature squeezing.

%============================================================================================================================================================================
\section{Quantum-noise free measurements}
\label{sec:sq-nfm}

In this section we will relate our operational quantification of nonclassicality to quantum noise effects in properly designed measurements.
We will start with the discussion of the well known scenarios of photon-number and quadrature measurements. Furthermore, we will show that a squeezed state can be used to realize noise-free quantum measurements in a particular measurement scenario. In this operational sense we may conclude that a squeezed state is perfectly nonclassical, even for a partial suppression of the quadrature vacuum noise.

Let us consider some physical consequences of perfect quantumness.
We may define perfect quantumness in a given experimental situation through Eq.~\eqref{eq:perf-qu-state}. A related  
observable, $\hat{f}^\dagger \hat{f}$, becomes a QNF variable. For this purpose it is sufficient to
prepare the system under study in a pure quantum state that fulfills the
Condition~(\ref{eq:perf-qu-state}). 
This opens possibilities to perform high-precision measurements at the ultimate limit of vanishing quantum noise.
Given an experimental setup and the related operator $\hat{f}$, one may solve Eq.~\eqref{eq:perf-qu-state} to derive optimal quantum
states for QNF measurements.

\subsection{Noise-free number statistics}

Consider the situation for the measurement of the photon number or the excitation number of a quantum mechanical harmonic oscillator. Let us consider the Jaynes-Cummings (JC) interaction~\cite{JCM} as it is realized in cavity QED~\cite{aamp20-347,prl54-551,prl58-353,prl76-1800}. In the vibronic motion of a laser-driven trapped ion the same Hamiltonian can be realized in the Lamb-Dicke regime~\cite{epl17-509}. In general, the dynamics is more rich and can be described by a nonlinear JC Hamiltonian~\cite{pra52-4214}, which has been confirmed in experiments~\cite{prl-76-1796}.

The standard JC interaction Hamiltonian is~\cite{JCM}
\begin{equation}
\label{eq:H-int-JC}
\hat{H}_{\rm int} = \frac{\hbar}{2} g  \hat{a} \hat{A}_{21} +  \mathrm{H.c.}\,,
\end{equation}
where $\hat{A}_{ij}=|i\rangle \langle j|$ ($i,j=1,2$) is the electronic flip 
operator. It describes the resonant interaction of an atomic two-level system with a quantum mechanical harmonic oscillator or a radiation mode, with coupling strength $g$. The system is prepared at $t=0$ in the state 
\begin{equation}
\label{eq:initial}
\hat{\varrho} (0) = |2\rangle \langle 2| \otimes \hat{\rho} (0),
\end{equation}
with $\hat{\varrho}$ and $\hat{\rho}$ denoting the full quantum state and that of the harmonic subsystem, respectively. The atom is observed in the excited state $|2\rangle$ with the probability
\begin{equation}
\label{eq:occup-JC}
p_{|2\rangle} (t)  = \frac{1}{2} \left \{ 1 +  {\rm Tr} \left [\hat{\rho} (0) \cos \left (g t \sqrt{\hat{a}^\dagger \hat{a} + 1} \right ) \right ]\right \}.
\end{equation}

For our problem we choose the operator $\hat f$ in Eq.~\eqref{eq:f-JC}.
According to Eq.~\eqref{eq:perf-qu-state-delta}, the optimal quantum state is
\begin{equation}
\label{eq:Fock}
\hat{a}^\dagger \hat{a} |n\rangle = n |n\rangle,
\end{equation}
with $n= \langle \hat{a}^\dagger \hat{a} \rangle$. This is the well-known Fock state, with a fixed excitation number of the harmonic subsystem. These states are clearly nonclassical according to the Condition~(\ref{eq:noncl-sub-no}), except the ground state, $n=0$. It is easily seen that the evolution
of $p_{|2\rangle} (t)$ according to Eq.~\eqref{eq:occup-JC} is a purely harmonic one in such a nonclassical state, the photon number being determined by the oscillation frequency. This is the typical situation for the QNF measurements under study. Any broadening of the number statistics leads to a non-harmonic dynamics, see the corresponding experiments~\cite{prl76-1800,prl-76-1796}.

\subsection{Noise-free quadrature statistics}

Let us briefly consider the situation of a noise-free quadrature measurement as it can be implemented for a properly laser-driven trapped ion~\cite{prl-75-2932,prl-94-153602}. This leads to a situation that resembles
that of the photon-number measurement based on the JC interaction. The required interaction Hamiltonian reads as~\cite{prl-75-2932}
\begin{equation}
\label{eq:H-int-quadr}
\hat{H}_{\rm int} = \hbar \Omega^\ast \hat{x}_\varphi \hat{A}_{21} +  \mathrm{H.c.}\,,
\end{equation}
where $\Omega$ is an effective Rabi frequency characterizing the laser excitation. For the physical realization we refer to the next subsection, where it follows as a special case of the generalized scheme studied there.

Let us consider the time evolution of the system starting again from the initial state~(\ref{eq:initial}). Now the electronic excitation evolves according to~\cite{prl-75-2932}
\begin{equation}
\label{eq:occup-quadr}
p_{|2\rangle} (t)  = \frac{1}{2} \left \{ 1 +  \int_{-\infty}^\infty dx \cos(x \tau)\, p(x,\varphi)\right \}.
\end{equation}
Here $\tau = |\Omega| t$ is the dimensionless time, $p(x,\varphi)={\rm Tr}[\hat{\rho} (0) |{x}_\varphi \rangle \langle {x}_\varphi|]$ 
is the probability distribution of the quadrature for arbitrary but fixed phase $\varphi$, and $|{x}_\varphi\rangle$ is the corresponding quadrature eigenstate.

Combining Eqs.~\eqref{eq:noncl-quadr}~and~\eqref{eq:perf-qu-state-delta}, the sought perfect quantum state is given by
\begin{equation}\label{eq:perf-sq-state}
\hat{x}_\varphi | \psi \rangle = {x}_\varphi | \psi \rangle\,,
\end{equation} 
which defines the quadrature eigenstate, $| \psi \rangle \equiv | {x}_\varphi
\rangle$, with the eigenvalue being ${x}_\varphi = \langle \hat{x}_\varphi
\rangle$. In this case the electronic dynamics according to 
Eq.~\eqref{eq:occup-quadr} is purely harmonic, as in the case of the JC dynamics for a Fock state. The oscillation frequency reflects again the precisely defined value of the observable to be measured.

This reproduces the well-known fact that the quadrature eigenstates are suited for QNF quadrature measurements. The severe
difficulty in realizing this situation consists in the fact that these
eigenstates represent the limit of infinitely strong squeezing, which requires an infinite amount of energy. Nevertheless, experimenters try to generate strongly squeezed states in order to approach this ideal situation. For example, recently a 10~dB reduction of the noise power of radiation has been achieved~\cite{prl-100-033602}, and even stronger squeezing was realized in the quantized motion of a trapped ion~\cite{prl-76-1796}. In this way one can suppress the noise effects in measurements significantly, but one cannot reach the QNF limit.

\subsection{Squeezed states for perfect measurements}

The question appears whether there is an alternative possibility of using squeezed states for QNF measurements. We will show that the answer is yes. It can be realized by a proper choice of the observable to be
measured. Since the quadrature noise is reduced for a squeezed states, it may seem that a quadrature measurement is the optimal choice. 
However, we may achieve a better performance and even
reach the QNF limit.

For simplicity, we will deal with a
squeezed vacuum state $|0;\nu \rangle$, which obeys the eigenvalue
equation
\begin{equation}\label{eq:sq-vac}
(\mu \hat{a} + \nu \hat{a}^\dagger) |0;\nu \rangle  = 0\,,  \quad \mu^2 -
|\nu|^2 =1\,.
\end{equation} 
The complex (real) parameter $\nu$ ($\mu$) controls the amount of quadrature noise reduction for properly fixed phase $\varphi$.  Total noise suppression requires $|\nu| \to \infty$, which cannot be realized in practice.
Instead, we may choose the operator $\hat{f}$ for our measurement device as
\begin{equation}\label{eq:sq-f}
\hat{f}\equiv \mu \hat{a} + \nu \hat{a}^\dagger.
\end{equation}  
By comparing Eq.~\eqref{eq:sq-vac}~with~Eq.~\eqref{eq:perf-qu-state}, it is obvious that the squeezed vacuum indeed obeys the condition of perfect quantumness for the resulting observable $\hat{f}^\dagger\hat{f}$. In this way one can  implement a QNF measurement.

\begin{figure}[h]
	\includegraphics[width=4.5cm]{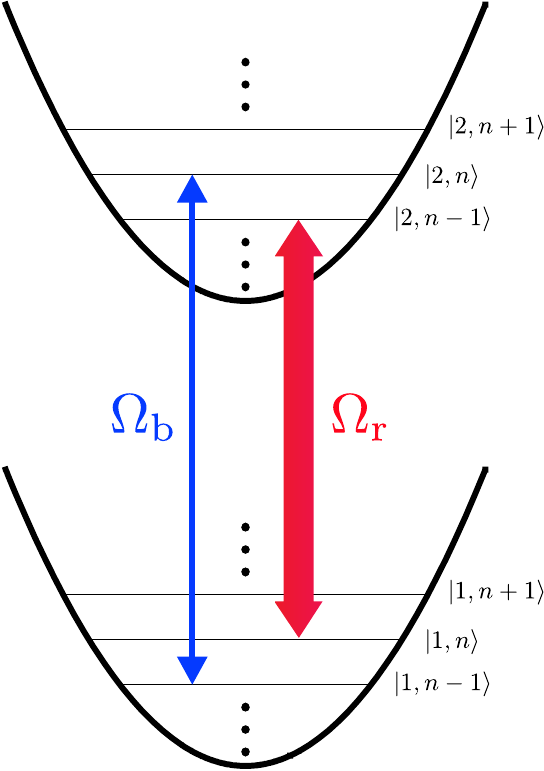}
	\caption{(color online). Scheme for a QNF measurement with moderate squeezing. The interaction on the red sideband is stronger than that on the blue one, $|\Omega_{\rm r}| > |\Omega_{\rm b}|$.}\label{fig:scheme}
\end{figure}

We still need a measurement setup for the
observable $\hat{f}^\dagger \hat{f}$.  
For this purpose we will consider the
situation for a trapped and laser-driven ion. In this case a motional squeezed state can be readily prepared~\cite{prl-76-1796}.
Let us consider a trapped ion, in the resolved-sideband and the Lamb-Dicke regimes. It is simultaneously laser-driven on the first red and blue sidebands, cf. Fig.~\ref{fig:scheme}. The couplings on the red and the blue  sidebands are given by the Rabi frequencies $\Omega_{\rm r}$ and $\Omega_{\rm b}$, respectively, which fulfill the condition $|\Omega_{\rm r}| > |\Omega_{\rm b}|$. For equal driving of both sidebands the scheme measures  quadratures as considered in the previous subsection. More details on the method, including the measurement of the time evolution of the electronic states of the atom, are given in~\cite{prl-75-2932,prl-94-153602}.

The interaction Hamiltonian reads as
\begin{equation}
\label{eq:H-int}
\hat{H}_{\rm int} = \frac{\hbar}{2} \hat{A}_{21} \mathrm{e}^{\mathrm{i}\varphi_{\rm r}}(|\Omega_{\mathrm{r}}|\hat{a}+ |\Omega_{\mathrm{b}}| \mathrm{e}^{\mathrm{i}\Delta \varphi} 
\hat{a}^\dagger) +  \mathrm{H.c.}\,,
\end{equation}
where
$\varphi_{\rm r}$ and $\Delta \varphi$ are the phase of the red-detuned laser and the phase difference of both lasers, respectively.
It can be rewritten as
\begin{equation}
\label{eq:H-int-f}
\hat{H}_{\rm int} = \frac{\hbar}{2} \Omega  \hat{f} \hat{A}_{21} +  \mathrm{H.c.}\,,
\end{equation}
with $\Omega= \mathrm{e}^{\mathrm{i}\varphi_{\rm r}} \sqrt{|\Omega_{\mathrm{r}}|^2 -|\Omega_{\mathrm{b}}|^2}$. The operator $\hat{f}$ is given by Eq.~\eqref{eq:sq-f}, with
\begin{equation}
\label{eq:nu}
\nu = \frac{|\Omega_{\mathrm{b}}|}{|\Omega|} 
\mathrm{e}^{\mathrm{i}\Delta \varphi},
\end{equation}
and $\mu^2 = 1+|\nu|^2$ according to Eq.~\eqref{eq:sq-vac}.
The resulting dynamics is sensitive to the operator $\hat{f}$ under study.

Straightforward algebra yields for the evolution of the electronic-state occupation of a trapped ion, initially prepared in the state~(\ref{eq:initial}), the result
\begin{equation}
\label{eq:occup}
p_{|2\rangle} (t)  = \frac{1}{2} \left \{ 1 +  {\rm Tr} \left [\hat{\rho} (0) \cos \left (|\Omega| t \sqrt{\hat{f}^\dagger \hat{f} + 1} \right ) \right ]\right \}.
\end{equation}
This evolution sensitively depends on the statistics of the observable $\hat{f}^\dagger \hat{f}$, with $\hat{f}$ from Eq.~\eqref{eq:sq-f}.
The ion is initially prepared in a motional squeezed vacuum according to Eq.~\eqref{eq:sq-vac}. In this case we easily arrive at
\begin{equation}
\label{eq:occup-coh}
p_{|2\rangle} (t)  = \frac{1}{2} \left [ 1 +   \cos \left (|\Omega| t \right )\right ].
\end{equation}
This represents a completely coherent oscillation, reflecting again the QNF property of the observable $\hat{f}^\dagger \hat{f}$. As in the examples considered above, the oscillation frequency 
yields the measured sharp expectation value of our basic observable, which is zero in the present case. 
The harmonic electronic dynamics clearly displays the striking 
property of the moderately squeezed states. For any amount of squeezing one may adjust the observable $\hat{f}^\dagger \hat{f}$ properly, such that the squeezed state exhibits perfect quantumness. The implemented detection scheme represents a perfect QNF measurement for this observable.

%============================================================================================================================================================================
\section{Algebraic quantification}
\label{sec:gen-quant}

We have seen above that very different states can be used for a perfect QNF measurement: the eigenstates of the Hermitian operators $\hat n$ and $\hat x_\varphi$ on one hand, and the eigenstates of the non-Hermitian operator $\hat f$ in Eq.~\eqref{eq:sq-f} on the other hand. As outlined above, the possibility of a QNF measurement is closely related to our concept of operational quantification of nonclassicality. Here we will consider the question of what are the common features of these very different quantum states to make them useful for perfect measurements. The answer to this question will open a new approach to the quantification of the nonclassicality of quantum states from the viewpoint of their algebraic structure.
For this purpose, we introduce the degree of nonclassicality.
The conclusions obtained from the algebraic quantification will be compared with those from the operational one. 

\subsection{Axiomatic quantification}

For the quantification of entanglement an axiomatic formulation of entanglement measures has been introduced~\cite{Measure1,Measure2,Measure3}.
Let us here formulate a similar approach for a nonclassicality measure.
Recently, a general method to introduce %an 
geometric ordering procedures and measures on convex sets of quantum states has been formulated~\cite{SpeVoOrdering}.

Before we can start with the definition of a nonclassicality measure, we need to consider mappings between quantum states.
Such linear functions can be given in terms of Kraus operators $\Lambda$, which map a quantum state $\hat\rho$ to another one $\Lambda(\hat\rho)$.
A Kraus operator is a classical one, if, by definition, it maps any classical state to (another) classical state:
\begin{eqnarray}
	\Lambda(|\alpha\rangle\langle\alpha|)=\int d^{2}\alpha'\, \Pi_{\alpha}(\alpha') |\alpha'\rangle\langle\alpha'|,
\end{eqnarray}
with $\Pi_{\alpha}$ being a classical probability distribution for all $\alpha\in\mathbb C$.
It is worth mentioning that it is sometimes convenient to relax this requirements to $\Pi_{\alpha}$ is non-negative.
The additionally needed normalization can be done via $\Lambda(\hat\rho)/{\rm tr}\,\Lambda(\hat\rho)$.
This class of Kraus operators refers to as nondeterministic Kraus operations.

Examples of classical Kraus operators can be given by $\Lambda_\beta(\hat\rho)=\hat D(\beta)\,\hat\rho\,\hat D(\beta)^\dagger$ (where $\hat D(\beta)$ denotes the unitary displacement operator), or the free time evolution, $\Lambda_\varphi(\hat\rho)={\rm e}^{-{\rm i}\varphi\hat n}\,\hat\rho\, {\rm e}^{{\rm i}\varphi\hat n}$ for $\varphi=\omega (t-t_0)$.
Obviously, both map a pure classical state -- i.e. a coherent state -- to another one.
More sophisticated is the transposition in Fock basis
\begin{eqnarray}
	\Lambda(\hat\rho)=\hat\rho^{\rm T},
\end{eqnarray}
which maps $|\alpha\rangle\langle\alpha|$ to $|\alpha^\ast\rangle\langle\alpha^\ast|$.
This map cannot be described as a unitary operator, but it also preserves the purity.

A non-purity preserving classical operation can be formulated, for example, as a convolution with a classical probability distribution $\Pi(\alpha)$,
\begin{eqnarray}
	\Lambda(|\alpha\rangle\langle\alpha|)
	=\int d^2\alpha'\,\Pi(\alpha'-\alpha)|\alpha'\rangle\langle\alpha'|.
\end{eqnarray}
Here, $\Pi$ introduces some classical noise to the coherent state $|\alpha\rangle$.
One example could be given by thermal noise, $\Pi(\alpha)=\frac{1}{\pi\bar n}{\rm e}^{-|\alpha|^2/\bar n}$, as it occurs in thermalization processes.
Alternatively, this particular map can be connected with the Lee measure~\cite{pra-44-R2775}.
A similar approach introduces some phase noise to a quantum state.
Such a classical phase-randomization has been experimentally realized to study its influence on the nonclassicality properties of squeezed light~\cite{Experiment5}.

Other realizations of such classical Kraus operators can be given by photon substraction experiments~\cite{Experiment2,Experiment3} and noiseless amplification~\cite{Experiment4,Theorie2}.
The photon substraction maps $P(\alpha)$ to $|\alpha|^2P(\alpha)$.
The noiseless amplification maps $|\alpha\rangle$ to $|g\alpha\rangle$, with $g>1$.
Both mappings belong to the class of nondeterministic (probabilistic) Kraus operators.
This means that the experiment, which realizes these Kraus operators, has a success rate below one and post-selection methods have to be applied.

After the definition of classical maps, let us now come to the definition of a proper nonclassicality measure.
A function $E_{\rm Ncl}$, mapping a quantum state to a non-negative number, which fulfills
\begin{enumerate}
	\item $E_{\rm Ncl}(\hat\rho)=0$, iff $\hat\rho$ classical;
	\item and for any classical Kraus operator $\Lambda$, we have
	\begin{eqnarray*}
		E_{\rm Ncl}(\hat\rho)\geq E_{\rm Ncl}(\Lambda(\hat\rho));
	\end{eqnarray*}
\end{enumerate}
refers to as a nonclassicality measure.
Note that such a measure is invariant under classical unitaries $\hat U$,
\begin{eqnarray}
	&E_{\rm Ncl}(\hat\rho)\geq E_{\rm Ncl}(\hat U\hat\rho\hat U^\dagger)\\
	\mathrm{ and }\nonumber
	&E_{\rm Ncl}(\hat U^\dagger[\hat U\hat\rho\hat U^\dagger]\hat U)\geq E_{\rm Ncl}(\hat\rho),
\end{eqnarray}
which yields $E_{\rm Ncl}(\hat\rho)=E_{\rm Ncl}(\hat\rho')$ for $\hat\rho'=\hat U\hat\rho\hat U^\dagger$.

\subsection{Degree of Nonclassicality}
Following the approach in Ref.~\cite{SpeVoOrdering}, we may construct such a nonclassicality measure.
First, we start with a measure for pure states, and, using a convex roof construction~\cite{ConvexRoof1,ConvexRoof2}, we will get a measure for mixed quantum states.
In order to quantify the quantum behavior of the system under study, we start with the quantum superposition of coherent states,
\begin{eqnarray}\label{Eq:Pure-r}
	|\psi_r\rangle=\sum_{k=1}^r \lambda_k |\alpha_k\rangle,
\end{eqnarray}
where $|\alpha_k\rangle$ are coherent states for $\alpha_k\neq\alpha_{k'}$ ($k\neq k'$), $\lambda_k$ are complex coefficients, and $r$ is a given integer.
Whenever a given pure state $|\psi_r\rangle$ can be written in such a form, we say it is in the set $\mathcal S_{\rm pure}^{(r)}$.

For mixed states, we define classical mixtures of those pure states as
\begin{eqnarray}\label{Eq:Mixed-r}
	\rho=\sum_{\psi_r\in\mathcal S_{\rm pure}^{(r)}} p_{\psi_r} |\psi_r\rangle\langle\psi_r|,
\end{eqnarray}
with $p_{\psi_r}\geq0$ and $\sum_{\psi_r} p_{\psi_r}=1$.
The set $\mathcal S^{(r)}$ is given by the closure (with respect to the trace norm) of states in the form of Eq.~\eqref{Eq:Mixed-r}.
We get a system of convex sets, which fulfill the inclusion $\mathcal S^{(r-1)}\subset\mathcal S^{(r)}$.
Finally, we can define the degree of nonclassicality $E_{\rm Ncl}(\hat\rho)$ as
\begin{eqnarray}
	E_{\rm Ncl}(\hat\rho)=1-{\rm e}^{-(r-1)},
\end{eqnarray}
where $r$ is the integer for which holds that $\hat\rho$ is in $\mathcal S^{(r)}$, but not in $\mathcal S^{(r-1)}$.

In case of a classical state $\hat\rho$, we have $\hat\rho\in\mathcal S^{(1)}$ and $\hat\rho\notin\mathcal S^{(0)}$ (the void set).
We get for classical states: $E_{\rm Ncl}(\hat\rho)=0$.
In contrast to the operational measure $R$, we have an 'if and only if' condition.
For $R=0$, we cannot conclude in general that the considered quantum state is a classical one.
In the operational sense, $R=0$ means that the considered measurement cannot use the possible nonclassicality within the state.

Let us consider perfect nonclassicality, $E_{\rm Ncl}(\hat \rho)=1$ or $r=\infty$.
This means that $\hat\rho\notin\mathcal S^{(r)}$, for finite $r$.
It is important to justify that $\hat\rho$ is indeed in the set $\mathcal S^{(\infty)}$.
Since any pure state can be written as an infinite superposition of coherent states,
any mixed state $\hat\rho$ can be formally written as a mixture of states with $r=\infty$.

Due to the fact that a classical operation $\Lambda$ maps the set $\mathcal S^{(r)}$ to a subset of itself,
%we obtain the result that 
the degree of nonclassicality $E_{\rm Ncl}$ fulfills the requirements of a nonclassicality measure.
Its physical interpretation is quite convenient.
By construction, the degree of nonclassicality identifies quantumness based on the quantum superposition principle.

\subsection{Quantum superpositions}

Let us compare the degree of nonclassicality with related entanglement measures.
The given measure is related to the Schmidt number~\cite{SchmidtNumber}, counting the number of superpositions of product states.
The Schmidt number is the convex roof construction of the Schmidt rank~\cite{SchmidtRank} of pure quantum states.
Recently it has been shown, that the Schmidt number is a universal entanglement measure~\cite{SpeVoUniversal}.
However, it only coincides with the degree of nonclassicality from the formal point of view.
Here, we quantify single mode nonclassicality of a harmonic oscillator.

The Schmidt number and the degree of nonclassicality consider quantumness by the quantum superposition principle, which yields all possible kinds of quantum interference phenomena.
The uncertainty principle is such a consequence.
Whenever the eigenvectors of a given observable $\hat A$ need to be written as a superposition of eigenvectors of another observable $\hat B$, those two observables do not commute, $[\hat A,\hat B]\neq 0$.
The result is that a simultaneous QNF measurement of $\hat A$ and $\hat B$ is impossible.
Note that this condition is necessary and sufficient.

For the quantum system of the harmonic oscillator, this also implies the well-known fact that the ground state has a non-zero energy.
The Hamiltonian may be written as $\hat H=c_1\hat p^2+c_2\hat x^2$ with two proper positive constants $c_1$, $c_2$, and $\hat p$ being the conjugate momentum of $\hat x$.
Due to the symmetry of the Hamiltonian for the exchanges $\hat x\mapsto -\hat x$ or $\hat p\mapsto -\hat p$, the minimal energy is given for a states with $\langle\hat x\rangle=\langle\hat p\rangle=0$.
Therefore, from $[\hat x,\hat p]\neq 0$ follows that
\begin{eqnarray}
	\langle \hat H\rangle_{\rm min}=c_1\langle (\Delta \hat p)^2\rangle+c_2\langle (\Delta \hat x)^2\rangle>0.
\end{eqnarray}
Hence the vacuum noise 
is also a consequence of the quantum superposition principle.

In addition, we may discuss some Kraus operators, which may increase the numer of superpositions of coherent states.
Obviously, such a Kraus operator is a nonclassical one, since the degree of nonclassicality can increase.
A possible generation of states with controlled number of superpositions $r$ has been studied for trapped ions~\cite{Theorie1}. In Experiments, generations of so-called Schr\"odinger cat states with $r=2$ have been realized, using trapped ions~\cite{Monroe}, Rydberg atoms~\cite{Noel}, cavity QED systems~\cite{Brune}, or propagating optical fields~\cite{Experiment1}.
Both methods consider Kraus operations which map $|\alpha\rangle$ to some superposition state, for example, $|\alpha\rangle+|-\alpha\rangle$.
Thus, these experiments can generate from a classical state with $r=1$ a nonclassical state with $r=2$.
Such experiments clearly use the quantum superposition principle to create this nonclassicality.
Another well-known Kraus operation which uses quantum properties is the photon addition, see e.g.~\cite{Experiment2}, which is given by the nondeterministic Kraus operator $\Lambda(\hat\rho)=\hat a^\dagger\hat\rho\hat a$.

It is not surprising that the quantum superposition principle leads to a manifold of quantum effects. 
It is remarkable that methods under study identify the number of superpositions $r$ -- defining $E_{\rm Ncl}$ -- as a proper nonclassicality measure.
Hence a fundamental concept of quantum physics directly defines a quantifier of quantumness.

\subsection{Examples}
In the previous examples of operational nonclassicality, we have shown that squeezed states and Fock states exhibit perfect operational nonclassicality in the sense of QNF measurements.
Let us study the amount of nonclassicality of these states using the proposed degree of nonclassicality, $E_{\rm Ncl}$.
In a first step, we consider properties of the nonclassicality number $r$ for pure states.
Afterward we apply our findings to Fock states and squeezed states.

Let us consider the resolution of the unity with coherent states,
\begin{eqnarray}
	\hat 1=\int_{\mathbb C} \frac{d^2\alpha}{\pi}\,|\alpha\rangle\langle\alpha|.
\end{eqnarray}
A quantum state $|\psi\rangle$ can be decomposed as
\begin{eqnarray}
	|\psi\rangle=\hat 1|\psi\rangle=\int_{\mathbb C} \frac{d^2\alpha}{\pi}\,{\rm e}^{-|\alpha|^2/2}\left({\rm e}^{|\alpha|^2/2}\langle\alpha|\psi\rangle\right)|\alpha\rangle.
\end{eqnarray}
In case we have a given nonclassicality number, $|\psi\rangle=|\psi_r\rangle\in\mathcal S_{\rm pure}^{(r)}$, we get
\begin{eqnarray}
	\hat 1|\psi_r\rangle=\int_{\mathbb C} \frac{d^2\alpha}{\pi}\,{\rm e}^{-|\alpha|^2/2}\left(\sum_{k=1}^r\lambda_k{\rm e}^{\alpha^\ast\alpha_k}{\rm e}^{-|\alpha_k|^2/2}\right)|\alpha\rangle.
\end{eqnarray}
This means that for each $\alpha$ must hold that
\begin{eqnarray}\label{Eq:Condition}
	{\rm e}^{|\alpha|^2/2}\langle\alpha|\psi\rangle=\sum_{k=1}^r\lambda_k{\rm e}^{\alpha^\ast\alpha_k}{\rm e}^{-|\alpha_k|^2/2}.
\end{eqnarray}
Hence a convenient test for the question {\it ''Is $|\psi\rangle$ in $\mathcal S^{(r)}$ or not?''} is given.
The state $|\psi\rangle$ is not in this set, iff it does not fulfill Eq.~\eqref{Eq:Condition}.

First, we may consider the photon Fock state $|n\rangle$ for $n>0$.
A photon represents the particle description of radiation.
A good quantifier should give a high nonclassicality of such a state.
We get from Eq.~\eqref{Eq:Condition}
\begin{eqnarray}
	\frac{{\alpha^\ast}^{n}}{\sqrt{n!}}=\sum_{k=1}^r\lambda_k{\rm e}^{\alpha^\ast\alpha_k}{\rm e}^{-|\alpha_k|^2/2}.
\end{eqnarray}
Except the constant polynomial, no other is a finite linear combination of exponential functions.
Here we have for any finite $r$ that ${\alpha^\ast}^n\notin{\rm span}\{{\rm e}^{\alpha^\ast\alpha_k}:k=1,\dots,r\}$ -- independently of the choice of the $\alpha_k$'s.
Thus we get $r=\infty$, which is equivalent to
\begin{eqnarray}
	E_{\rm Ncl}(|n\rangle\langle n|)=1.
\end{eqnarray}
Photons exhibit perfect nonclassicality.
Beyond the operational usefulness for the considered QNF measurement, the algebraic nonclassicality is also maximal.

The second example is a squeezed state, whose nonclassicality arises from the sub-vacuum noise in one quadrature.
The squeezed vacuum state $|0;\nu\rangle$ can be written as
\begin{eqnarray}
	|0;\nu\rangle=\frac{1}{\sqrt\mu}{\rm e}^{-\frac{\nu}{2\mu}{\hat{a}^\dagger}{}^{2}}|0\rangle,
\end{eqnarray}
with a squeezing parameter $\xi\neq0$ ($\mu=\cosh|\xi|$ and $\nu=\sinh|\xi|{\rm e}^{{\rm i}\arg(\xi)}$)~\cite{book}.
Now, Eq.~\eqref{Eq:Condition} reads as
\begin{eqnarray}
	\frac{1}{\sqrt{\mu}}{\rm e}^{-\frac{\nu}{2\mu}{\alpha^\ast}{}^{2}}=\sum_{k=1}^r\lambda_k{\rm e}^{\alpha^\ast\alpha_k}{\rm e}^{-|\alpha_k|^2/2}.
\end{eqnarray}
Again, one can argue that a function with a quadratic exponent cannot be written as a finite combination of functions with a linear exponent.
Immediately, we have $r=\infty$ or
\begin{eqnarray}
	E_{\rm Ncl}(|0;\nu\rangle\langle0;\nu|)=1.
\end{eqnarray}
Squeezed states exhibit maximal nonclassicality, not only in the previously discussed operational sense, but also in the algebraic approach of nonclassicality measures.
This underlines the previously obtained result from the relative operational nonclassicality $R$.
It comes as a surprise that the squeezed state is perfectly nonclassical, without considering the limit of infinite squeezing.
Both approaches -- the operational and the algebraic one -- manifest the fact that sub-vacuum noise is a very strong kind of nonclassicality.
It is worth noting that a strong nonclassicality is easy to identify, e.g. by low-order moments, cf. Section~\ref{sec:establ-ex}.

%============================================================================================================================================================================
\section{Summary and conclusions}
\label{sec:sum}

In conclusion we have introduced an operational measure for the nonclassicality or quantumness of a quantum state of the harmonic oscillator.
It is based on the negativity of an observable whose classical counterpart is positive semi-definite.
The resulting perfect quantumness is related to the feasibility of performing totally quantum-noise-free measurements.
As an example, we have demonstrated that a moderately squeezed state of the quantized center-of-mass motion of a trapped ion can display perfect quantumness.
An implementation of the corresponding noise-free quantum measurement has been given.
In this context we have also discussed the general strategy of implementing quantum-noise-free measurements.

A general, axiomatic quantification of nonclassicality has been introduced and compared with the operational approach.
For this purpose, the degree of nonclassicality has been defined.
This degree is zero for classical states, and it becomes one for maximal nonclassicality.
We have outlined the relation between fundamental concepts -- namely the quantum superposition principle -- in quantum optics and the defined degree of nonclassicality.

Examples of maximally nonclassical states have been discussed.
A Fock state represents the particle aspects of a quantized electromagnetic wave.
This dualism is reflected by the maximal degree of nonclassicality for  photon number states, from the viewpoint of classical electrodynamics.
A squeezed state, on the other hand, can beat the classical shot-noise limit of a quadrature measurement.
The violation of such an ultimate classical boundary is again verified by the maximal degree of nonclassicality even for moderate squeezing. It is important to stress,
that in practice stronger squeezing may be preferred, whenever it is sufficiently robust against occurring classical noise effects. 

Both concepts, the operational and the algebraic one, present different ways for the quantification of nonclassicality.
The operational approach can quantify the nonclassicality, which can be measured or used in a particular experimental scenario. Experimental noise effects could be included in the choice of the observable.
The algebraic degree of nonclassicality quantifies the actual amount of quantumness independently of the experiment.
The first method is of great interest for particular applications of nonclassical light.
The second method identifies -- for an experimentally generated or theoretically studied state -- how much nonclassicality can be possibly used.

%============================================================================================================================================================================
\vspace*{1ex}
\section*{Acknowledgments}
The authors gratefully acknowledge valuable comments by C. Di Fidio, and support by the Deutsche Forschungsgemeinschaft through SFB 652.

\end{document}